\documentstyle[preprint,epsfig,aps]{revtex}

\begin{document}

\draft
%\twocolumn[\hsize\textwidth\columnwidth\hsize\csname@twocolumnfalse%
%\endcsname
%\baselineskip 18pt
%\draft

\title{Dispersion Relation Analysis of Neutral Pion Photo- and Electroproduction
 at Threshold using the MAID and SAID solutions}
\author{S. S.  Kamalov\cite{Sabit}, L. Tiator, D. Drechsel}
\address{Institut f\"ur Kernphysik, Universit\"at Mainz, 55099 Mainz,
Germany}
\author{R. A. Arndt, C. Bennhold, I. I. Strakovsky and R. L. Workman}
\address{Center for Nuclear Studies, Department of Physics, \\
The George Washington University, Washington D.C. 20052, U.S.A. }

\date{\today}
\maketitle

\begin{abstract}
Neutral pion photo- and electroproduction at threshold is analyzed
in the framework of dispersion relations. For this purpose, we
evaluate the real threshold amplitudes in terms of Born
contributions and dispersion integrals determined by the imaginary
parts of the MAID and SAID multipoles. The results show
considerable cancellations between Born terms and resonance
contributions. Good agreement with the data is found for
photoproduction.  While our dispersion analysis suggests
considerable discrepancies for electroproduction, the present
state of the experimental multipole analysis at finite $Q^2$ does
not permit drawing conclusions at this time.
\end{abstract}
%]
%\pacs{PACS numbers: 13.60.Le, 14.20.GK \\
% KEYWORDS: pion, photoproduction, electroproduction, Low Energy Theorem,
% threshold }

\section{Introduction}
Electro- and photoproduction of neutral pions near threshold have
been a topic of many experimental and theoretical investigations
over the past decade. Triggered by surprising results obtained at
Saclay~\cite{Sac}, the Mainz~\cite{Mai} and
Saskatoon~\cite{Bergstrom} groups established that a formerly
believed low-energy theorem (LET) ~\cite{Let1,Let2} for S-wave
photoproduction was at variance with nature. While the LET
predicted a threshold S-wave multipole
$E_{0^+}=-2.4\cdot10^{-3}/m_{\pi}$, the experiment yielded
$E_{0^+}\approx-1.3\cdot10^{-3}/m_{\pi}$. The discrepancy between
the theorem and the experimental data was finally explained by
Bernard {\it et al.}~\cite{Ber} who showed that loop corrections
provided nonanalytical terms in the pion mass $\mu$. The flaw of
the low-energy theorem was therefore the assumption that the
amplitudes would be an analytical function in the pion mass $\mu$,
which could be expanded in a Taylor series in the soft-pion limit.
In the following years, these calculations were considerably
refined by evaluating the S-wave amplitude $E_{0^+}$ to order
$p^4$ in the chiral expansion, and the 3 P-wave amplitudes
($E_{1^+},\ M_{1^+},\ {\rm and}\ M_{1^-}$) up to order $p^3$.
While there appear 3 low-energy constants to that order, two
combinations of P-wave amplitudes were found to be independent of
these constants.  Further work has extended this approach to
virtual photons~\cite{Bernard}

Using a different approach, a recent calculation obtained a good
description of $\pi^0$ photo- and electroproduction in the
threshold region within a meson-exchange dynamical
model~\cite{KY99,DMT}. It was found that the largest contributions
to the final-state interaction came from one-loop charge-exchange
rescattering. This approach lead to a to good description of the
$S$-wave multipoles.

The large reduction of the S-wave threshold amplitude was
independently obtained using fixed-$t$ dispersion
relations~\cite{HDT}. In this approach, the Born terms have to be
evaluated at the nucleon pole where the pseudovector and the
pseudoscalar pion-nucleon coupling are identical. While the result
of the old LET was essentially equivalent to the result of
pseudovector coupling at threshold, the value of the multipole at
the pole position corresponds to pseudoscalar coupling. As a
result the Born term  to be used in dispersion theory is $E_{0^+}
(\rm{pole}) = -7.6\cdot10^{-3}/m_{\pi}$, and thus the dispersion
integrals over the excited states have to cancel about 80~\% of
the pole term in order to describe the data.

In Ref.~\cite{HDT}, the coupled-integral equations were solved
using the method of Omn\`{e}s and Mushkashevili~\cite{omnes}. On
the condition that the complex phases of the multipoles are known
and with given assumptions for their high-energy behavior, this
method allows one to find unique solutions. In practice, however,
the phases are known only in the energy region below the two-pion
threshold due to the Watson theorem~\cite{FW}. Extending these
calculations to energies above the second resonance region, which
coincides with the onset of two-pion production, requires modeling
the phases by functions which depend on the pion-nucleon phase
shifts and inelasticity parameters. The ansatz for the functional
dependence is based on unitarity but by no means unique, and in
principle has to be determined by a fit to the data. It is
therefore the aim of the present work to extend the energy range
of the dispersion analysis by use of the Unitary Isobar
Model~\cite{MAID98} (called MAID in the following) as an input for
the imaginary parts of the multipole amplitudes. At the same time,
we want to compare the results obtained by use of MAID with those
with the SAID multipoles~\cite{SAID}. This allows us to present a
qualitative ``error band'' for the dispersion analysis, which
often has been asked for.

Our paper is organized as follows. In Section II, we briefly
recall the ingredients of dispersion relations at fixed $t$. The
actual calculations are described in Section III. In particular,
we extend the energy range of the MAID model by including the
contributions from all $S-,\ P-,\ D-,$ and $F-$wave resonances
with four-star PDG status. As a particularly sensitive test of the
extended model, we present predictions of our calculation for
threshold production of neutral pions in Section IV.

\section{Dispersion relations for pion electroproduction}

In the present work, we will use fixed-$t$ dispersion relations
(DR) to construct the pion electroproduction multipoles (or
partial waves) ${\cal {\widetilde M}}$,
\begin{eqnarray}
Re\,{\cal {\widetilde M} }_{\alpha}(W,Q^2)& = & {\cal {\widetilde
M}}_{\alpha}^{Pole}(W,Q^2)+
\frac{P}{\pi}\int^{\infty}_{W_{thr.}}dW'\,\frac{Im\, {\cal
{\widetilde M}}_{\alpha}(W',Q^2)}{W'-W}
\nonumber\\
& + &
\frac{1}{\pi}\int^{\infty}_{W_{thr.}}dW'\sum_{\beta}\,
{\cal {\widetilde K}}_{\alpha\beta}(W,W',Q^2)\, Im\, {\cal
{\widetilde M}}_{\beta}(W',Q^2) \,,
\label{dr1}
\end{eqnarray}
where $\alpha$ and $\beta$ are the set of quantum numbers, $W$ is
the total c.m. energy of the $\pi N$ system, and $Q^2={\bf
k}^2-\omega^2 >0$ is the four-momentum squared of the virtual
photon with three-momentum ${\bf k}$ and energy $\omega$. The
first term in Eq.~(\ref{dr1}), ${\cal {\widetilde
M}}_{\alpha}^{Pole}$, comprises the explicitly known contributions
from the pole diagrams with pseudoscalar $\pi NN$ coupling. The
second and third terms are the principal value and regular parts
of the dispersion integrals which contain the kernels ${\cal
{\widetilde K}}_{\alpha\beta}$ and the imaginary parts of the
multipoles. Both integrals run only over the physical region
starting at threshold $W_{thr}=m+\mu$, where $m$ and $\mu$ are the
nucleon and pion masses, respectively.

The detailed expressions for the kernels and the numerical recipes
for their numerical computation are given in Ref. \cite{Gehlen}.
In accordance with this work, the relations between the multipoles
${\cal {\widetilde M}}_{\alpha} =({\cal {\widetilde
E}}_{l\pm},\,{\cal {\widetilde M}}_{l\pm}, \,{\cal {\widetilde
L}}_{l\pm}/\omega)$ and the standard CGLN~\cite{CGLN} multipoles
$(E_{l\pm},\,M_{l\pm},\,L_{l\pm})$ are the following:
\begin{eqnarray}
\label{mult}
& &{\cal {\widetilde E}}_{l+}=
-8\pi\frac{\sqrt{{\cal E}_1/{\cal E}_2}}{(qk)^l k^2} E_{l+}\,,
\quad\qquad
{\cal {\widetilde E}}_{l+1,-}=
-8\pi\frac{\sqrt{{\cal E}_2/{\cal E}_1}k^2 W}{(qk)^{l+1}} E_{l+1,-}\,,
\nonumber\\
& &{\cal {\widetilde M}}_{l+}= 8\pi\frac{\sqrt{{\cal E}_1/{\cal
E}_2}W}{(qk)^l } M_{l+}\,,\qquad
{\cal {\widetilde M}}_{l+1,-}=
-8\pi\frac{\sqrt{{\cal E}_2/{\cal E}_1}}{(qk)^{l+1}} M_{l+1,-}\,,\\
& &{\cal {\widetilde L}}_{l+}=
8\pi\frac{\sqrt{{\cal E}_1/{\cal E}_2}}{(qk)^l k^2 } L_{l+}\,,
\qquad\qquad
{\cal {\widetilde L}}_{l+1,-}=
8\pi\frac{\sqrt{{\cal E}_2/{\cal E}_1}}{(qk)^{l+1}} L_{l+1,-}\,,
\nonumber
\end{eqnarray}
with ${\cal E}_{1(2)}=E_{1(2)}+m$, where $E_{1(2)}$ denotes the
nucleon c.m. energy in the initial (final) state, $q=\mid {\bf q}
\mid$ and $k=\mid {\bf k} \mid$ the absolute values of the c.m.
pion and photon momenta, respectively, and $l$ the pion orbital
momentum.

While the fixed-$t$ DR in the form of Eq.~(\ref{dr1}) are uniquely
defined, the separation into the principal value and regular
integral contributions is not unique and depends on the choice of
the kinematical factors in Eq.~(\ref{mult}). Other kinematical
factors, i.e., as used in Refs.~\cite{BDW,Rollnik,HDT}, will
change the relative contributions of these two integrals and the
expressions for the kernels. For example, if we introduce a new
set of multipoles via the relation ${\cal {\widetilde
M'}}_{\alpha}(W)={\cal {\widetilde M}}_{\alpha}(W)/f_{\alpha}(W)$
with a certain factor $f_{\alpha}(W)$, we find the following
relation between the new and old kernels:
\begin{eqnarray}
{\cal {\widetilde
K'}}_{\alpha\beta}(W,W')=\frac{f_{\beta}(W')}{f_{\alpha}(W)}\,
{\cal {\widetilde K}}_{\alpha\beta}(W,W')+
\delta_{\alpha\beta}\,\frac{f_{\beta}(W')-f_{\alpha}(W)}{f_{\alpha}(W)(W'-W)}\
. \label{kern}
\end{eqnarray}
The different expressions for the kernels given in the literature
can be easily checked and compared by use of these relations. For
example, we found that at $Q^2=0$, the kernels from
Ref.~\cite{Gehlen} and Ref.~\cite{BDW} lead to the same result.

For future analysis, it is convenient to  rewrite the DR of
Eq.~(\ref{dr1}) in terms of the CGLN multipoles ${\cal
M}_{\alpha}=(E_{l\pm},\,M_{l\pm},\,L_{l\pm}/\omega)$
\begin{eqnarray}
Re\,{\cal M}_{\alpha}(W)= {\cal M}_{\alpha}^{Pole}(W)+ {\cal M}_{\alpha}^{Diag}(W)+
\frac{1}{\pi}\int^{\infty}_{W_{thr.}}dW'\sum_{\alpha\neq \beta}\,K_{\alpha\beta}(W,W')\,
Im\, {\cal M}_{\beta}(W') \,,
\label{dr2a}
\end{eqnarray}
where
\begin{eqnarray}
{\cal
M}_{\alpha}^{Diag}(W)=\frac{P}{\pi}\int^{\infty}_{W_{thr.}}dW'
\frac{Im {\cal M}_{\alpha}(W')r_{\alpha}(W')}{(W'-W)
r_{\alpha}(W)}+
\frac{1}{\pi}\int^{\infty}_{W_{thr.}}dW'K_{\alpha\alpha}(W,W')\,
Im {\cal  M}_{\alpha}(W').
\label{dr2b}
\end{eqnarray}
The kinematical factor $r_{\alpha}(W)$ is determined by
Eq.~(\ref{mult}) with the relation ${\cal {\widetilde
M}}(W)=r_{\alpha}(W)\,{\cal M}(W)$, and
$K_{\alpha\beta}(W,W')={\cal {\widetilde
K}}_{\alpha\beta}(W,W')\,r_{\beta}(W')/r_{\alpha}(W)$. One of the
advantages of such a representation is that each term in
Eq.~(\ref{dr2a}) is individually independent of the choice for the
kinematical factor $r_{\alpha}$. This statement can be easily
proved by use of Eq.~(\ref{kern}).

\section{Calculations of the dispersion integrals}

One of the methods widely used to calculate the dispersion
integrals in Eq.~(\ref{dr1}) or Eqs.~(\ref{dr2a})-(\ref{dr2b}) is
based on the Watson theorem~\cite{FW}, stating that the phase of
pion photo- and electroproduction is equal to the phase shift of
pion-nucleon scattering, $\delta_{\alpha}(W)$, below the two-pion
threshold. Below this threshold, we can therefore use the
following relation between the real and imaginary parts of the
amplitude:
\begin{eqnarray}
Im \, {\cal M}_{\alpha}(W,Q^2)= Re \,{\cal M}_{\alpha}(W,Q^2)\
\tan{\delta_{\alpha}(W)}\,.
\label{Watson}
\end{eqnarray}
If we further make an assumption about the high-energy behavior of
the multipole phases, we obtain a system of coupled integral
equations for $Re {\cal M}_{\alpha}(W)$. This is the standard
method to apply fixed-$t$ dispersion relations to pion
photoproduction at threshold and in the $\Delta(1232)$ resonance
region, which was successfully used by many
authors~\cite{HDT,BDW,Rollnik,aznaurian}. The reliability of this
method at low energies ($W<1400$ MeV) is mainly based on the
finding that Eq.~(\ref{Watson}) can be applied to the important
$P_{33}$ multipole, dominated by the $\Delta(1232)$ resonance
contribution, with good accuracy up to $W=1600$ MeV.

Another method to calculate the dispersion integrals is based on
isobaric models~\cite{Salin,Loub,Walecka,Crawford} which allow
extending the use of fixed-$t$ DR  to higher energies. With this
approach, the imaginary parts of the pion photo- and
electroproduction multipoles are expressed in terms of background
(${\cal M}^B$) and resonance (${\cal M}^R$) contributions,
\begin{eqnarray}
 Im \, {\cal M}_{\alpha}(W,Q^2)= Im \, {\cal M}^B_{\alpha}(W,Q^2)
 + Im \, {\cal M}^R_{\alpha}(W,Q^2).
\end{eqnarray}
In the present work, both parts will be modeled similar to the
recently developed Unitary Isobar Model~\cite{MAID98} MAID. The
imaginary parts from the background appear due to final-state
interaction effects for the pions produced by nonresonant
mechanisms and contain contributions from both the Born terms
($V^{Born}_{\alpha}$) with an energy-dependent mixing of
pseudovector-pseudoscalar (PV-PS) $\pi NN$ coupling and t-channel
vector-meson exchanges ($V^{\omega,\rho}_{\alpha}$),
\begin{eqnarray}
{\cal M}^B_{\alpha}(W,Q^2)= [V^{Born}_{\alpha}(W,Q^2)+
V^{\omega,\rho}_{\alpha}(W,Q^2)]\,(1+iT^{\alpha}_{\pi N}(W))\,,
\end{eqnarray}
where the pion-nucleon scattering amplitude $T^{\alpha}_{\pi N}=
\frac{1}{2i}[\eta_{\alpha}\exp{(2i\delta_{\alpha})} - 1]$ is given
in terms of the $\pi N$ phase shifts $\delta_{\alpha}$ and the
inelasticity parameters $\eta_{\alpha}$, taken from the analysis
of the SAID group\cite{VPI97}. In accordance with
Ref.~\cite{MAID98} the background contribution depends on 5
parameters: The PV-PS mixing parameter $\Lambda_m$ in
$V^{Born}$(see Eq.~(12) of Ref.\cite{MAID98}) and 4 coupling
constants in $V^{\omega,\rho}$. Note that in our present work, we
do not include hadronic form factors at the $\omega NN$ and $\rho
NN$ vertices.

Following Ref.~\cite{MAID98} the resonance contributions are given
in terms of Breit-Wigner amplitudes,
\begin{equation}
{\cal M}^R_{\alpha}(W,Q^2)\,=\,{\bar{\cal
A}}_{\alpha}^R(Q^2)\, \frac{f_{\gamma R}(W)\Gamma_R\,M_R\,f_{\pi
R}(W)}{M_R^2-W^2-iM_R\Gamma_R} \,e^{i\phi_R}\,,
\label{eq:BW}
\end{equation}
where $f_{\pi R}$ is the usual Breit-Wigner factor describing the
decay of a resonance $R$ with total width $\Gamma_{R}(W)$ and
physical mass $M_R$. The main parameters in the resonance
contributions are the strengths of the electromagnetic transitions
described by the reduced amplitudes ${\bar{\cal
A}}_{\alpha}^R(Q^2)$, which have to be extracted from the analysis
of the experimental data. In the present work, we extend the
previously developed MAID model by including contributions from
all $S-,\,P-,\,D-$ and $F$-wave resonances with four-star PDG
status~\cite{PDG}. The addition of new resonances requires
performing a new fit. For this purpose, we use the SAID data
base~\cite{Base} for pion photoproduction in the energy range
$W_{thr}<W<2000$~MeV with 15,700 data points. The resonance
parameters and values of $Im {\cal M}$ at the resonance position
obtained from the best fit are listed in Table~1. We note that in
most cases the background contributions to the imaginary parts are
less then 10\% at the resonance positions. The only exceptions are
the channels with the $S_{31}(1620)$, $S_{11}(1650)$, and
$P_{31}(1910)$ resonances, for which we find $ Im {\cal M}^B=$
2.50, 0.85 and -1.45, respectively, in comparison to $ Im {\cal
M}^R=$ -1.28, 2.45 and 0.52.  Here and in the following, all
multipoles are quoted in units of $10^{-3}/m_{\pi+})$. In the case
of the overlapping resonances in the $S_{11}$ proton channel, we
find $Im\,_pE^{(1/2)}(1535)=3.32+0.14+0.37$ and
$Im\,_pE^{(1/2)}(1650)=0.43+1.17+0.85$, where the first and second
terms are the contributions from the first and second $S_{11}$
resonance, respectively, and the last terms come from the
background contributions.

Alternatively, we calculate the dispersion integrals using the
solution SM02 of the SAID multipole analysis~\cite{SAID} (see
Table 1). Concerning the integration up to infinity, we assume
that the multipoles have an asymptotic behavior like $1/W$ for
$W\ge2300$~MeV. This is the minimal power providing convergence
for the GDH sum rules~\cite{GDH}. In the threshold region, we
introduce the pion mass difference by assuming that the imaginary
part of the $E_{0+}$ multipoles is proportional to the $\pi^+$
momentum below $W=1090$~MeV. This assumption is based on the fact
that near threshold the main contribution to the imaginary part
comes from the coupling with the $\pi^+ n$ channel~\cite{DMT}.

\section{Results and discussions}

\subsection {$\pi^0$ photoproduction at threshold}

The threshold region has traditionally posed a problem to the
analysis of $\pi^0$ photoproduction within a dispersion-relation
approach~\cite{BDW}. This is due mainly to considerable
cancellations in the dispersion integrals of Eqs.~(\ref{dr2a}) and
(\ref{dr2b}). As shown in Ref.~\cite{HDT}, by solving the integral
equations using the Watson theorem, the real part of the
$E_{0+}(\pi^0 p)$ threshold multipole obtains surprisingly large
contributions from the imaginary parts of higher multipoles which
peak at much larger energies. As a result, the high-energy region
provides sufficiently large contributions to nearly cancel the
nucleon pole term with pseudoscalar $\pi NN$ coupling, thus
leading to agreement with the experimental threshold values.

Similar results are obtained in our present work using fixed-$t$
DR and imaginary parts of the multipoles taken from the MAID model
and from the results of the SAID multipole analysis,
\begin{equation}
E_{0+}^{thr}(p\pi^0)=-7.89 + 2.84 + 4.09 -0.48-0.25+0.40 =
-1.29\qquad {\rm DR(MAID)}\,,
\label{E0MAID}
\end{equation}
\begin{equation}
E_{0+}^{thr}(p\pi^0)=-7.89 + 2.83 + 4.23 -0.51-0.14 +0.13 =
-1.35\qquad {\rm DR(SAID)}\,,
\label{E0SAID}
\end{equation}
where the contributions on the right-hand side are presented, in
accordance with Eq.~(\ref{dr2a}), in the following order: the pole
term, the diagonal $E_{0+}$, the kernel terms $M_{1+}$, $M_{1-}$,
$E_{1+}$,  and the combined kernel contributions of the higher
$D$- and $F$-wave multipoles. According to Eq.~(\ref{dr2b}), the
diagonal $E_{0+}$ contribution can be further divided into the
principal-value integral and the regular integral, which
contribute $1.23 + 1.61$ using MAID and $1.31 + 1.52$ using SAID
solutions. As discussed above, this sum does not depend on the
choice for the kinematical factor $r_{\alpha}(W)$. The individual
contributions from the coupling to the $D$- and $F$- wave
multipoles are presented in Table 2. Taken separately, they are
not negligible, but in the sum they nearly cancel and lead to a
total value very close to the extracted value of
Ref.~\cite{Schmidt}, $E_{0+}^{thr}(p\pi^0) = -1.33 \pm 0.11$.

Fig.~1 compares the energy dependence of the $E_{0+}$ amplitude,
obtained, on the one hand, directly from the MAID and SAID
solutions (dash-dotted curves) and, on the other hand, using of
the dispersion relations, Eq.~(\ref{dr1}), with $Im {\cal M}$ as
input taken from the MAID and SAID solutions (solid curves). We
clearly see the Wigner cusp effect appearing in the DR solutions
due to the infinite derivative of $Im E_{0+}$ (dashed curves) at
the charged pion threshold. In the MAID solution (dash-dotted
curve), the cusp effect is the result of the strong coupling to
the $\pi^+$ channel taken into account by the K-matrix
approximation~\cite{DMT}. The SAID solution does not include this
effect.

Finally, Table~3 summarizes our results for the threshold $S$- and
$P$-wave multipoles and compares them to the results of the recent
experimental analysis of Ref.~\cite{Schmidt}. For the $P$-wave
multipoles we list the values of the following linear
combinations,
$P_1=3E_{1+}+M_{1+}-M_{1-},\,P_2=3E_{1+}-M_{1+}+M_{1-}$ and
$P_3=2M_{1+}+M_{1-}$. In general, the DR results are consistent
with the corresponding MAID or SAID solutions and in good
agreement with the results of ChPT and the experimental values of
Ref.~\cite{Schmidt}. A large discrepancy remains for the $P_3$
amplitude, where the theoretical predictions with and without the
use of DR are considerably smaller than the experimental value.
This may hint at problems in the description of the $M_{1-}$
multipole which appears more pronounced in $P_3$ than in $P_1$ and
$P_2$.

\subsection {$\pi^0$ electroproduction at threshold}

Dispersion relations for pion electroproduction are more involved
due to the more complicated structure of the kernels
$K_{\alpha\beta}(W,W',Q^2)$. In addition, the transverse
multipoles of the virtual photons are also coupled with the
longitudinal ones via the kernels. Moreover, we have very limited
information about the longitudinal (Coulomb) resonance excitations
at finite $Q^2$. In the following, we present first calculations
for threshold $\pi^0$ electroproduction using dispersion relations
with the  dispersion integrals determined by the MAID model. The
longitudinal excitation of the $\Delta(1232)$ and $P_{11}(1440)$
resonances are described as shown in Ref.~\cite{MAID98}. For the
other resonances we assume the validity of the pseudothreshold
relation~\cite{DT} $E_{l\pm}^R=\pm \frac{k}{2\omega}(2j+1)
L_{l\pm}^R$. These assumptions lead to the following threshold
values for the $S$-wave multipoles at $Q^2=0.1$ (GeV/c)$^2$:
\begin{equation}
E_{0+}^{thr}(p\pi^0)=-3.69 + 2.46 + 2.96 -0.08 = 1.55\qquad {\rm
DR(MAID)}\,,
\label{E0thr}
\end{equation}
\begin{equation}
L_{0+}^{thr}(p\pi^0)=-3.76 + 0.54 + 1.82 +0.01 = -1.41\qquad {\rm
DR(MAID)}\,.
\label{L0thr}
\end{equation}
The terms on the right-hand side correspond, in that order, to the
contributions of the pole term, the diagonal term, the coupling to
the $M_{1+}$, and the coupling to the higher multipoles. As in the
case of real photons, we find that the largest contributions come
from the diagonal term and the $M_{1+}$ multipole, which nearly
cancel the large contribution of the pole term.

The threshold behavior of the $E_{0+}$ and $L_{0+}$ multipoles at
$Q^2=0.1$ (GeV/c)$^2$ is shown in Fig.~2. We point out the much
smaller cusp effect in the $L_{0+}$, compared to the $E_{0+}$
multipole, due to the smaller imaginary part of the $L_{0+}$.  The
fixed-$t$ DR results are in a good agreement with the results of
the analysis of Ref.~\cite{Distler}. On the other hand, the real
parts of the $E_{0+}$ and $L_{0+}$ multipoles obtained from the
MAID solution, are closer to the results of
Refs.~\cite{DMT,NIKHEF}. However, as discussed in Refs.~\cite{DMT}
and \cite{Merkel}, the extracted results for the $S$ waves at
finite $Q^2$ strongly depend on the assumptions used for the
$P$-wave contributions. This is especially true for the $E_{0+}$
multipole. For example, at $Q^2=0.1$ (GeV/c)$^2$ the differences
in the $P$ waves used by various groups lead to quite different
threshold values for the $E_{0+}$, namely $1.96\pm
0.33$~\cite{Distler}, $2.28\pm 0.36$~\cite{DMT}, and $0.58 \pm
0.18$~\cite{Merkel}. Clearly, these differences in the analysis
techniques must be resolved before a comparison with theoretical
predictions can be meaningful. Note that we find significant
dispersion corrections for both multipoles at finite $Q^2$.

Fig. 3 shows the $Q^2$ dependence for several $S$-wave multipoles
and $P$-wave multipole combinations and compares our results with
the results of the analyses of Refs.~\cite{Merkel,Distler}. A
number of interesting features emerge. In general, the DR results
for the transverse multipoles are consistent with the
corresponding MAID solution. For the $L_{0+}$ multipole, and the
longitudinal P-wave combinations $P_4$ and $P_5$, strong
dispersion corrections appear at low $Q^2$. Our dispersion results
are in agreement with the results from ChPT below $Q^2 < 0.05
GeV^2$ in the case of the $E_{0+}$ multipole and the $P_1$
combination but differ significantly for the $L_{0+}$ multipole,
and the $P_{23}^2=(P_2^2+P_3^2)/2$, $P_4=4L_{1+}+L_{1-}$ and
$P_5=L_{1-}-2L_{1+}$ amplitudes. This may reflect the fact that
some of the ChPT low-energy constants where fitted to
electroproduction threshold data while the MAID solutions are
constrained by data in the resonance sector. Just as in Fig.2, the
experimental points shown have to be understood in the context of
model-dependent analyses techniques.

Finally, we present in Fig. 4 predictions for the quantity $\Delta
P_{23}^2=(P_2^2 - P_3^2)/2$ which determines the sign of the beam
asymmetry, i.e. $\Sigma \sim -\Delta P_{23}^2$.  Recent
measurements~\cite{Schmidt} yielded a negative value for the
$\Delta P_{23}^2$ at $Q^2=0$ and $E_{\gamma}=160$ MeV, in rough
agreement with ChPT results. However, both the MAID and the DR
results are positive at the photon point and become more positive
for higher photon virtualities.  In contrast, the ChPT results
remain negative.  Clearly, a measurement of this observable at
finite $Q^2$ is highly desirable.

\section{Conclusion}

Threshold pion photo- and electroproduction have been calculated
with fixed-t dispersion relations. Unlike previous work for
photoproduction following the method of Omnes and Mushkashevili,
we have used the imaginary parts of the multipoles of the unitary
isobar model MAID and the phenomenological partial-wave analysis
SAID as input to calculate dispersion integrals.

Unitarity, crossing symmetry, Lorentz invariance and gauge
invariance are all fulfilled by the dispersion relations.
Especially crossing symmetry can only be partially fulfilled in
model calculations, even field-theoretical lagrangians violate
crossing symmetry when energy-dependent widths for nucleon
resonances are introduced. Rather than fitting to threshold data,
by using the dispersion relations we employ models that are fitted
to data in the resonance region, where more data is available.

For pion photoproduction we obtain very good agreement with the
threshold multipoles obtained from experimental analyses. Both the
cusp effect and pion-loop effects are well described and the
differences between the MAID and SAID inputs play only a minor
role. In fact, it rather reveals the systematic uncertainty in
such a dispersion approach. We also find good agreement with the
results of ChPT for for s- and p- waves, except for the quantity
$P_2^2 - P_3^2$. This discrepancy was already observed in the
previous dispersion analysis of Hanstein et al.\cite{HDT} and
relates to a very delicate cancellation among two large p-wave
amplitudes.

The situation for pion electroproduction reflects much
uncertainty, both in theory and experiment. Much less data is
available which leads to model dependencies in the extraction of
the multipoles at finite $Q^2$. Since the electroproduction
coincidence cross section cannot be completely separated, a model
independent analysis as in the photoproduction case is not yet
possible, making any comparison with theory difficult. We
emphasize that our dispersion theoretical calculation has the
advantage that most of the input for the fixed-t dispersion
relation comes from the magnetic excitation of the $\Delta$
resonance which is very well known even for pion
electroproduction.  Future experiments will hopefully remove the
model dependencies in the extraction of the multipole amplitudes
and allow an unambiguous comparison with the predictions from
dispersion relations.

\acknowledgements The Mainz group acknowledges support from the
Deutsche Forschungsgemeinschaft (SFB 443) and from the joint
Germany-Russia Heisenberg-Landau project.

The SAID group (RAA, IIS, and RLW) acknowledges partial support
from the U.~S.~Department of Energy Grant DE--FG02--99ER41110 and
from Jefferson Lab by the Southeastern Universities Research
Association under DOE contract DE--AC05--84ER40150. C.B.
acknowledges support from the U.~S.~Department of Energy Grant
DE--FG02--95ER--40907. C.B., L.T. and S.S.K. were also supported
by a NATO Collaborative Research Grant.

S.S.K are grateful to the Department of Physics at The George
Washington University for the hospitality extended during his
visit.

\begin{table}[htbp]
\begin{center}
\begin{tabular}{|c|ccc|cc|lc|}
&  &  & & MAID    &   & SAID  &  \\
$N^*$  &$M_R$[MeV]&
$\Gamma_R$[MeV]& $\beta_{\pi}$ & ${\cal M}_E$ & ${\cal M}_M$&
${\cal M}_E$ & ${\cal M}_M$\\
\hline $P_{33}(1232)$ & 1232 & 130 &
1.0  & -0.81  & 36.85 & -0.54  & 36.01 \\
 $P_{11}(1440)$ & 1440 &
350 & 0.70 &  ---   &  2.75 &  ---   &  2.74 \\
$D_{13}(1520)$ &
1520 & 130 & 0.60 &  4.56  &  1.97 &  5.31  &  2.18 \\
$S_{11}(1535)$ & 1520 & 80& 0.40 & 3.83& ---  &  3.77  &
---  \\
$S_{31}(1620)$ & 1620 & 150 & 0.25 & -1.28   & ---  &
-0.79 &  ---  \\
 $S_{11}(1650)$ & 1690 & 100 & 0.85 &   2.45  &
---  &   3.81 &  ---  \\
$D_{15}(1675)$ & 1675 & 150 & 0.45 &
0.10  &  0.32 &  0.03  &  0.25 \\
$F_{15}(1680)$ & 1680 & 135 &
0.70 &  1.77  &  1.23 &  1.80  &  1.20 \\
 $D_{33}(1700)$ & 1740 &
450 & 0.15 & -3.54  &  0.25 & -2.83  &  0.72 \\
 $P_{13}(1720)$ &
1720 & 250 & 0.20 &  0.55  & -0.07 &  0.58  &  0.02 \\
$F_{35}(1905)$ & 1905 & 350 & 0.10 &  0.45  &  0.32 &  0.40  &
0.29 \\
$P_{31}(1910)$ & 1910 & 200 & 0.25 &  ---   &  0.52 &  ---
&  0.83 \\
 $F_{37}(1950)$ & 1950 & 300 & 0.20 &  0.02  &  1.45 &
0.04  &  1.36 \\
\end{tabular}
\end{center}
\caption{Model parameters of the nucleon resonances in the proton
channels (resonance mass $M_R$, width $\Gamma_R$, pion branching
ratio $\beta_{\pi}$) and corresponding resonance + background
values of the imaginary parts of the electric $({\mathcal{M}}_E)$
and magnetic multipoles $({\mathcal{M}}_M)$ at resonance (in units
of $10^{-3}/m_{\pi+})$
obtained with the MAID2002 and SAID(SM02) solutions. % $^{*)}$
The partial branching ratios for the $S_{11}(1535)$ are assumed to be
$\beta_{\pi}=0.40,\ \beta_{\eta}=0.50,\ {\rm and}\
\beta_{2\pi}=0.10$. }
\end{table}

\begin{table}[htbp]
\begin{center}
\begin{tabular}{|c|cccccccc|}
  & $E_{2-}$ & $M_{2-}$ &  $E_{2+}$ & $M_{2+}$
  & $E_{3-}$ & $M_{3-}$ &  $E_{3+}$ & $M_{3+}$\\
\hline
 MAID & 0.16 &  --0.16  & 0.06  & 0.04 & 0.34 & --0.37 & --0.01 & 0.34  \\
 SAID & 0.28 &  --0.51  & 0.05  & 0.07 & 0.37 & --0.51 & --0.02 & 0.40  \\
\end{tabular}
\end{center}
\caption{Individual contributions of the $D$- and $F$- multipoles (in
units of $10^{-3}/m_{\pi+}$) to the multipole $E_{0+}^{thr}(\pi^0
p)$ at threshold. }
\end{table}

\begin{table}[htbp]
\begin{center}
\begin{tabular}{|c|cccc|}
 solutions  & $E_{0+}$ &  $P_1$ &  $P_2$ & $P_3$ \\
\hline
MAID2002 & --1.23 &  9.07  & --10.68  & 7.07  \\
DR(MAID) & --1.29 &  9.64  & --10.29  & 8.22  \\
\hline
SAID SM02& ----   & 8.79  & --11.23  & 9.60  \\
DR(SAID) & --1.35 &  9.70  & --10.46  & 8.91  \\
\hline analysis   & $-1.33 \pm 0.11 $ & $ 9.47 \pm 0.33 $ & $-9.46 \pm
0.39$
& $ 11.48 \pm 0.41 $ \\
\end{tabular}
\end{center}
\caption{ $E_{0+}$ (in units of $10^{-3}/m_{\pi^+}$) and
$P_1,\,P_2$ and $P_3$ (in units of $10^{-3}q/m^2_{\pi^+}$) for
photoproduction at threshold. The values extracted from the data
are taken from the analysis of Ref.\protect\cite{Schmidt}.}
\end{table}

%%%%%%%%%%%%%%   Fig. 1 ****************
\begin{figure}[h]
\centerline{\epsfig{file=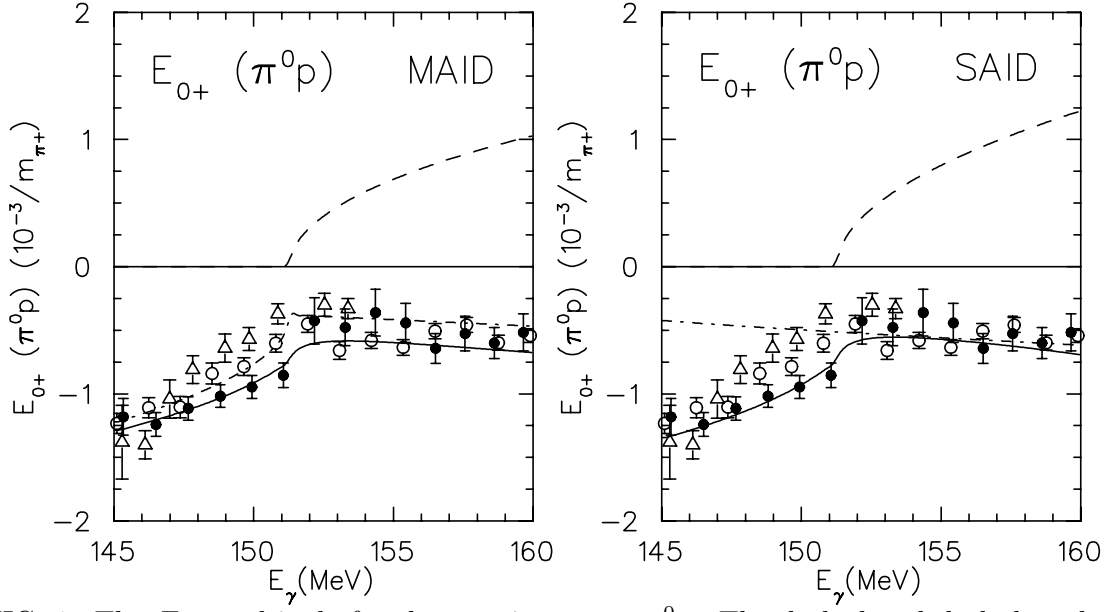, width=8 cm, angle=90} }
\caption{The $E_{0+}$ multipole for the reaction $\gamma
p\rightarrow \pi^0 p$. The dashed and dash-dotted curves show the
imaginary and real parts, respectively, as obtained from the
MAID2002 (left panel) and SAID solution SM02 with a modified
imaginary part as explained in the text (right panel). The solid
curves are the predictions for the real parts obtained with the
dispersion relations. The data points are the result of the
multipole analyses from Ref. \protect\cite{Fuchs96}($\triangle $),
Ref. \protect\cite{Bergstrom}($\bullet$), and Ref.
\protect\cite{Schmidt}($\circ$).}
\end{figure}

\newpage
%%%%%%%%%%%%%%   Fig. 2 ****************
\begin{figure}[h]
\centerline{\epsfig{file=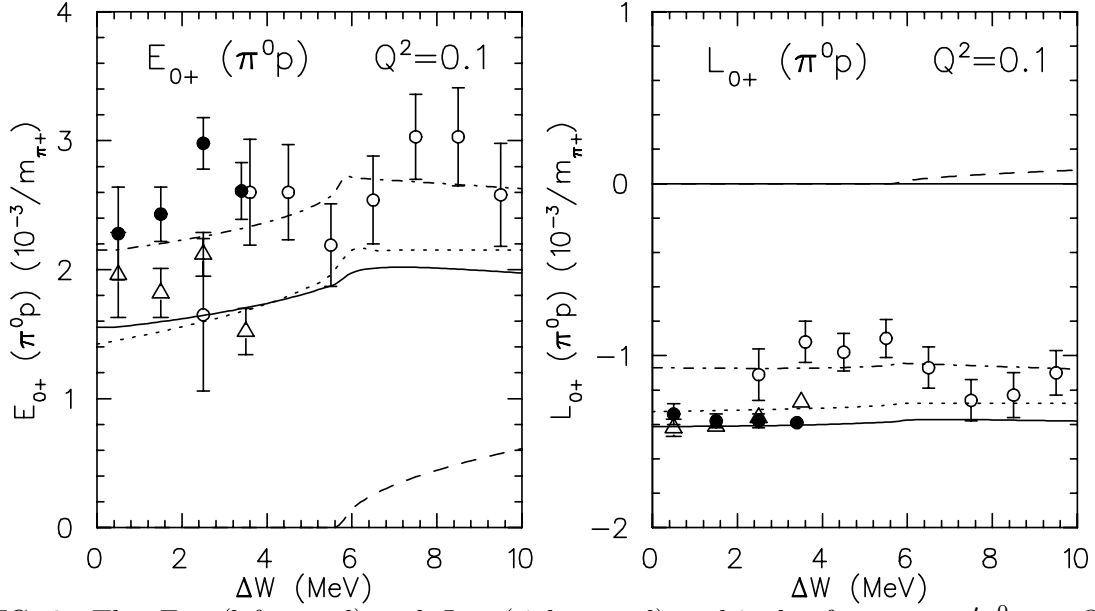, width=8 cm, angle=90} }
\caption{The $E_{0+}$ (left panel) and  $L_{0+}$ (right panel)
multipoles for $ep \rightarrow e'\pi^0 p$ at $Q^2$=0.1 (GeV/c)$^2$
as a function of $\Delta W= W-W_{thr}$. The dashed and dash-dotted
curves are the imaginary and real parts, respectively, for the the
MAID2002 solution. The solid curves are the predictions for the
real parts obtained with the dispersion relations. The data points
are the result of the analyses from
Ref.\protect\cite{NIKHEF}($\circ$),
Ref.\protect\cite{Distler}($\triangle$) and
Ref.\protect\cite{DMT}($\bullet$).}
\end{figure}

\newpage
%%%%%%%%%%%%%%   Fig. 3 ****************
\begin{figure}[h]
\centerline{\epsfig{file=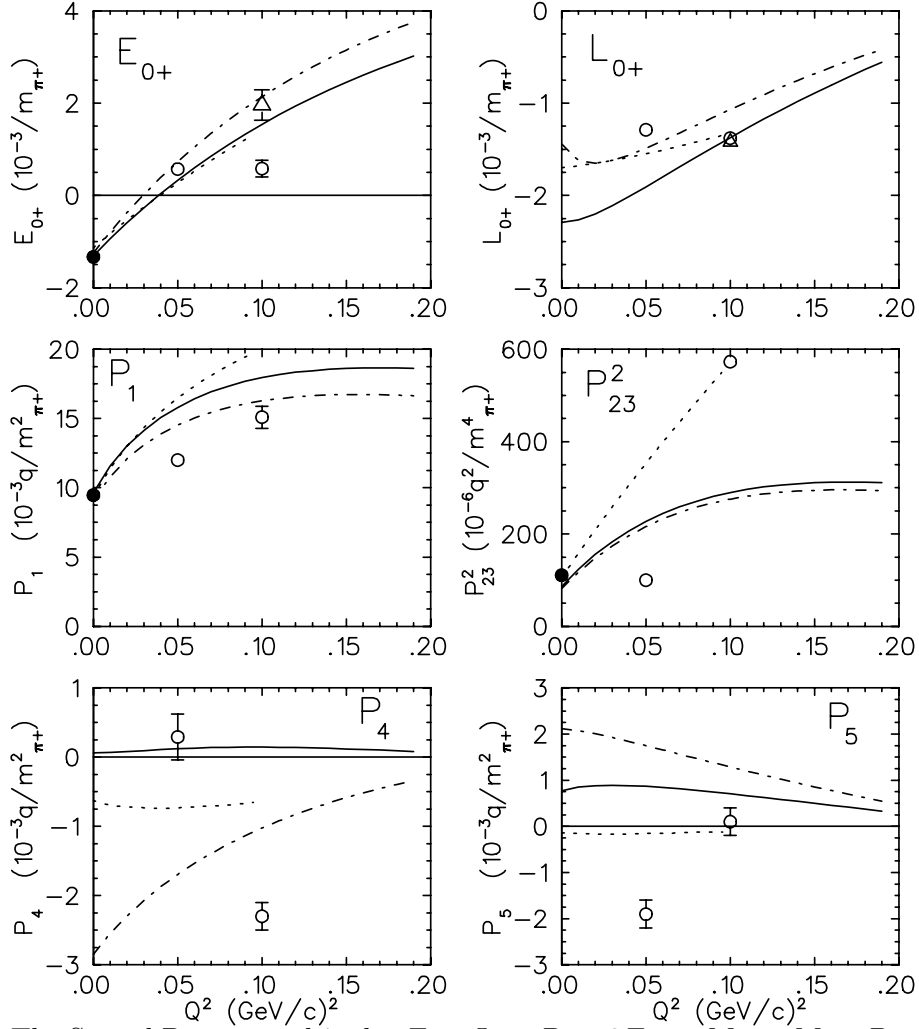, width=12 cm, angle=0} }
\caption{ The S-- and P--wave multipoles $E_{0+}$, $L_{0+}$,
$P_1=3E_{1+}+M_{1+}-M_{1-}$, $P_4=4L_{1+}+L_{1-}$,
$P_5=L_{1-}-2L_{1+}$, and $P_{23}^2=(P_2^2+P_3^2)/2$ for the
reaction $ep\rightarrow e'\pi^0 p$ at threshold as a function of
$Q^2$. The dash-dotted and solid curves are the MAID2002 solution
and the prediction of dispersion relations, respectively. The
dotted curves show the results of ChPT~\protect\cite{Bernard}. The
data points are the results of the analyses from
Ref.\protect\cite{Distler}($\triangle$)
Ref.\protect\cite{Merkel}($\circ$) and
Ref.\protect\cite{Schmidt}($\bullet$).}
\end{figure}

\newpage
%%%%%%%%%%%%%%   Fig. 4 ****************
\begin{figure}[h]
\centerline{\epsfig{file=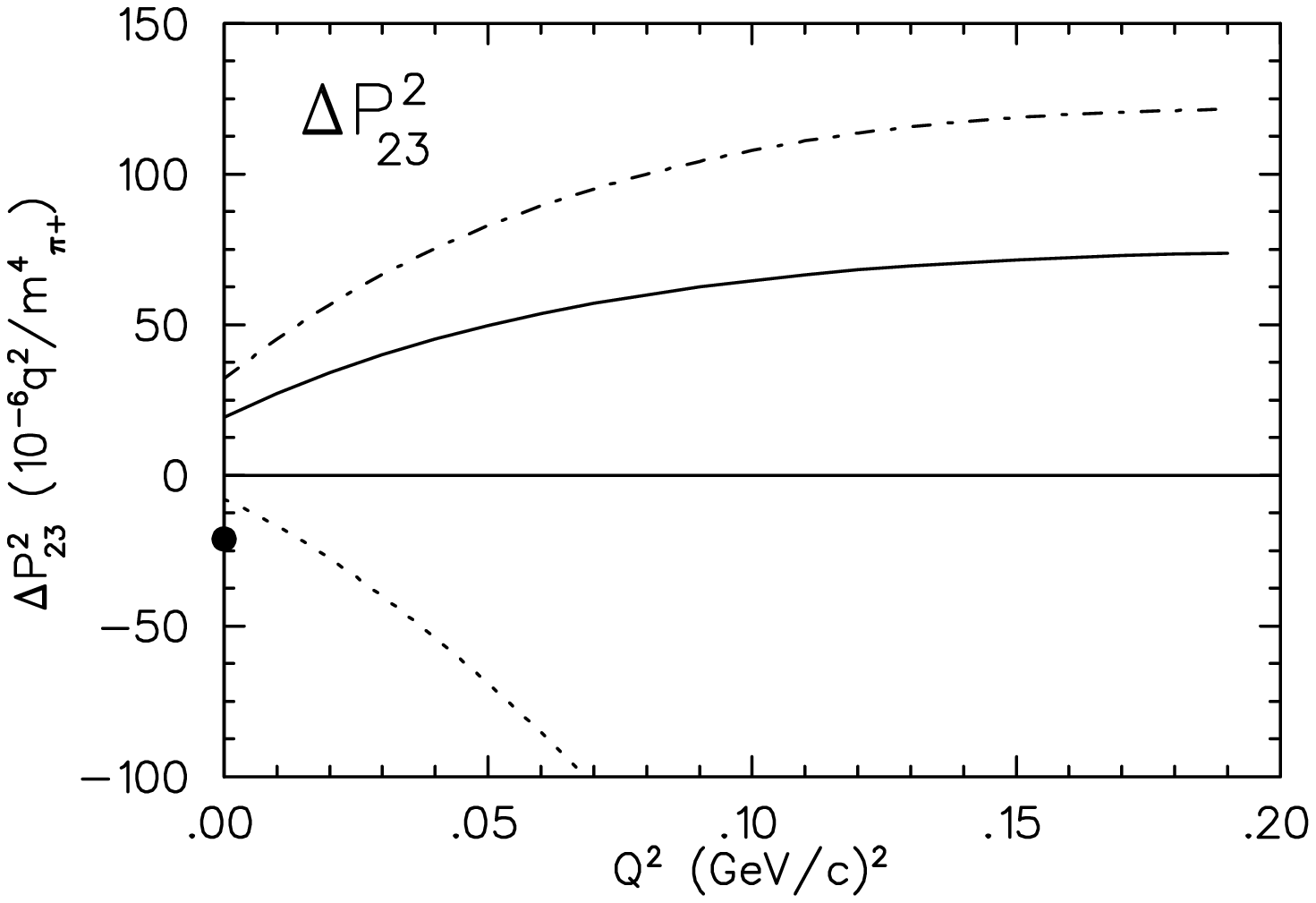, width=7 cm, angle=90} }
\caption{$\Delta P_{23}^2=(P_2^2-P_3^2)/2$ for the reaction
$ep\rightarrow e'\pi^0 p$ at threshold as a function of $Q^2$. The
notation of the curves is as in Fig. 3. The data point at $Q^2=0$
is the result of the analysis from Ref.\protect\cite{Schmidt}.}
\end{figure}


\begin{references}
\bibitem[*]{Sabit}
Permanent address: Laboratory of Theoretical Physics, JINR Dubna,
141980 Moscow region, Russia.
\bibitem{Sac}    E.~Mazzucato {\it et al.}, Phys. Rev. Lett.
                 {\bf 57}, 3144 (1986).
\bibitem{Mai}    R.~Beck {\it et al.}, Phys. Rev. Lett.
                 {\bf 65}, 1841 (1990).
\bibitem{Bergstrom}J.~C.~Bergstrom \textit{et al.} Phys.\ Rev.\ C\
                  \textbf{53}, 1052 (1996), {\it ibid} C \textbf{ 55},
                  2016 (1997).
\bibitem{Let1}   P.de~Baenst, Nucl. Phys. B {\bf 24}, 633 (1970).
\bibitem{Let2}   I.~A. Vainshtein and V.~I.~Zaharov,
                 Nucl. Phys. B {\bf 36}, 589 (1972).
\bibitem{Ber}    V.~Bernard, J.~Gasser, N.~Kaiser and U.-G.~Mei{\ss}ner,
                 Phys. Lett. B {\bf 268}, 219 (1991).
\bibitem{Bernard} V.~Bernard, N.~Kaiser, and Ulf-G.~Mei{\ss}ner,
                   Z. Phys. C {\bf 70}, 483 (1996); Nucl. Phys.
                   {\bf A607}, 379 (1996),{\bf A633}, 695 (1998) (E);
         and references contained therein.
\bibitem{KY99}   S.~S.~Kamalov and S.~N.~Yang, Phys.\ Rev. \ Lett.\
                 \textbf{83}, 4494 (1999).
\bibitem{DMT}    S.~S.~Kamalov, G.~Y.~Chen, S.~N.~Yang, D.~Drechsel, and
                 L.~Tiator, Phys.\ Lett.\ \textbf{B522}, 27 (2001).
\bibitem{HDT}    O.~Hanstein, D.~Drechsel, and L.~Tiator,
                 Nucl.\ Phys.\ \textbf{A632}, 561 (1998).
\bibitem{omnes}  R.~Omn\`{e}s, Nuovo Cim.{\bf 8}, 316 (1958).
\bibitem{FW}     K.~M.~Watson, Phys. Rev. {\bf 95}, 228 (1954)
\bibitem{MAID98} D.~Drechsel, O.~Hanstein, S.~S.~Kamalov, and L.~Tiator,
                 Nucl.\ Phys.\ \textbf{A645}, 145 (1999).
\bibitem{SAID}   R.~A.~Arndt, I.~I.~Strakovsky, and R.~L.~Workman,
                 submitted to Phys. Rev. C, Eprint nucl-th/0205067.
\bibitem{Gehlen} G.~v.~Gehlen, Nucl.\ Phys.\ \textbf{B9}, 17 (1968).
\bibitem{CGLN}   G.~F.~Chew, M.~L.~Goldberger, F.~E.~Low, and Y.~Nambu,
                 Phys.\ Rev.\ \textbf{106}, 1345 (1957).
\bibitem{BDW}    F.~A.~Berends, A.~Donnachie, and D.~L.~Weaver,
                 Nucl.\ Phys.\ \textbf{B4}, 1 (1967).
\bibitem{Rollnik}D.~Schwela, H.~Rollnik, R.~Weizel, and W.~Korth,
                 Z.\ f\"ur Phys.\ \textbf{202}, 452 (1967).
\bibitem{aznaurian} I.~G.~Aznauryan, Phys.\ Rev.\ D \ \textbf{57}, 2727 (1998).
\bibitem{Salin}  Ph.~Salin, Nuovo Cimento, \ \textbf{32}, 521 (1964).
\bibitem{Loub}   J.~P.~Loubaton, Nuovo Cimento, \textbf{39}, 591 (1965).
\bibitem{Walecka}J.~D.~Walecka, Phys.\ Rev.\ \textbf{162}, 1462 (1967).
\bibitem{Crawford}R.~L.~Crawford and W.~T.~Morton, Nucl.\
                 Phys.\ \textbf{B211}, 1 (1983); Particle
                 Data Group, Phys.\ Lett.\ \textbf{B239}, 1 (1990),
                 R.~L.~Crawford, in: {\it {Proceedings of
                 NSTAR2001, Mainz, Germany, March 7--10,
                 2001}}, Eds. D.~Drechsel and L.~Tiator,
                 World Scientific, p.~163.
\bibitem{VPI97}  R.~A.~Arndt, I.~I.~Strakovsky, R.~L.~Workman, and M.M. Pavan,
                 Phys.\ Rev.\ C\ \textbf{52}, 2120 (1995).
\bibitem{PDG}    D.~E.~Groom \textit{et al.}, {\it {Review
                 of Particle Physics}}, Eur.\ Phys.\ J.\ C\
                 \textbf{15}, 1 (2000).
\bibitem{Base}   R.~A.~Arndt, I.~I.~Strakovsky, and R.~L.~Workman, in preparation,
                 SAID photoproduction database is available via
                 {\it http://gwdac.phys.gwu.edu}.
\bibitem{GDH}    S.~B.~Gerasimov, Yad. Fiz. \textbf{2}, 598 (1965)
                 [ Sov. J. Nucl. Phys. {\bf 2}, 430 (1966)];
                 S.~D.~Drell and A.~C.~Hearn, Phys.\ Rev.\
                 Lett.\ \textbf{16}, 908 (1966).
\bibitem{Schmidt}A.~Schmidt \textit{et al.} Phys.\ Rev.\ Lett.\
                 \textbf{87}, 232501 (2001).
\bibitem{Fuchs96}M.~Fuchs \textit{et al.} Phys.\ Lett.\ \textbf{B368},
                 20 (1996).
\bibitem{DT}     D.~Drechsel and L.~Tiator, J.\ Phys.\ G:\ Nucl.\ Phys.\
                 \textbf{18}, 449 (1992).
\bibitem{Merkel} H.~Merkel \textit{et al.} Phys.\ Rev.\ Lett.\
                 \textbf{88}, 012301 (2002).
\bibitem{Distler}M.~O.~Distler \textit{et al.} Phys.\ Rev.\ Lett.\
                 \textbf{80}, 2294 (1998).
\bibitem{NIKHEF} H.~B.~van~den~Brink \textit{et al.} Phys.\ Rev.\ Lett.\
                 \textbf{74}, 3561 (1995), Nucl.\ Phys.\ \textbf{A612},
                 391 (1997).

\end{references}
\end{document}